\documentclass[preprint,showpacs]{revtex4}
\usepackage[dvips]{graphicx}
\usepackage{amsmath,amsfonts,amssymb,bm,ulem}

\begin{document}
\title{Interference of guiding polariton mode in ``traffic'' circle waveguides composed
of dielectric spherical particles}

\author{I. Ya. Polishchuk}
\affiliation{Department of Chemistry, Tulane University, New
Orleans, LA 70118, USA} 
\affiliation{RRC Kurchatov Institute, Kurchatov Sq., 1, 123182 Moscow,
Russia}
\affiliation{Max-Planck-Institut f\"{u}r Physik Komplexer Systeme, D-01187}

\author{M. I. Gozman}
\affiliation{Department of Chemistry, Tulane University, New
Orleans, LA 70118, USA} 
\affiliation{RRC Kurchatov Institute, Kurchatov Sq., 1, 123182 Moscow,
Russia}

\author{G. S. Blaustein}
\affiliation{Department of Chemistry, Tulane University, New
Orleans, LA 70118, USA} 

\author{A. L. Burin}
\affiliation{Department of Chemistry, Tulane University, New
Orleans, LA 70118, USA}

\begin{abstract}
The interference of polariton guiding modes propagating through "traffic circle" waveguides composed of
dielectric spherical particles is investigated. The dependence of intensity of the wave on the position of the 
particle was studied using the multisphere the Mie scattering formalism. We show that if the frequency of light
belongs to the passband of the waveguide, electromagnetic waves may be considered as two optical beams running along a circle in opposite
directions and interfering with each other. Indeed, the obtained intensity behavior can be represented as 
a simple superposition of two waves propagating around a circle in opposite directions. The applications of this interference are discussed. 
\end{abstract}

\maketitle

\section{Introduction}

Low-dimensional aggregates of nanoparticles absorbing, scattering and guiding electromagnetic waves are attracting growing interest
because they can be used in various micro- and nano-systems where optical energy is received, transferred and converted on a subwavelength scale
\cite{Astratov, Deych, Shore,Fan,Meade,Tang-Kotov,Mayer1,Mayer2,Burin CircArr,Burin LowDimArr,Me ICTON2007,Me OptExpress}. 
If the material's refractive index is sufficiently large, then these aggregates can possess long-lived quasistates, which are polariton modes 
representing the superposition of light and polarization oscillation of the material \cite{Burin CircArr,Me OptExpress}. 
These modes propagate within the aggregates, while their losses due to light emission are very small.  Consequently, low-dimensional aggregates of 
particles can be used as waveguides for optical energy transport. This transferred energy can stimulate a number of processes such as
the photoexcitation of a molecule or a chemical reaction\cite{Mayer1,Mayer2}.

For example, if we excite the polariton mode within a linear chain of particles using a point harmonic source located near the end of the chain, 
then there exists a frequency domain, i. e. the passband, where almost all of the energy from the source will be guided by the chain \cite{Me PhysLett}.  
Other modes in the related frequency domain can be treated as bound or guiding modes. To our knowledge, the first nano-waveguide capable of transferring 
optical energy a distance of 100 nm was constructed by Atwater and
coworkers \cite{Mayer1,Mayer2} as a linear chain of spherical silver or gold
particles. The use of metallic particles has an advantage in that the losses due to photon emission can be almost completely 
suppressed \cite{Me OptExpress}, yet the efficiency of energy transfer
is suppressed due to optical energy absorption by free electrons.

Guiding systems of dielectric particles have negligible absorption losses, while radiative losses can also be avoided as dielectric materials have characteristically larger 
refractive indices with $n_{r}>2$ \cite{Burin CircArr,Me OptExpress}. The finite size of the
guiding energy band in periodic arrays of particles results in the the formation of slow light modes at the band edge \cite{Me OptExpress}. 
Therefore, it may be more convenient to use dielectric materials
having very weak light absorption\ \cite{Burin CircArr,Burin
LowDimArr,Me ICTON2007,Me OptExpress}. Frequency pass bands of
linear chains were investigated \cite{Me ICTON2007, Me
OptExpress, Me ICTON2008, Me PhysLett}and it was found that the top
of the frequency pass band corresponds to the top of the Brillouin
band edge in quasi-momentum space, while the bottom of the
frequency pass band is determined by the guiding mode criterion (see
Eq.(\ref{CondBand}) below) \cite{Burin CircArr, Me OptExpress}. It
was discovered that the closer the frequency of electromagnetic
waves emitted by the source to the upper boundary of the pass band
and the longer the chain, the more effective the waveguide 
and the lower the amount energy is dissipated into the surroundings. It was shown that
for $N\rightarrow +\infty $, the rate of decay of electromagnetic
waves decreases as $\gamma (N)\sim N^{-3}$ \cite{Burin
LowDimArr,Me ICTON2007,Me OptExpress}.

Another example of a low-dimensional aggregate of particles possessing
wave-guiding properties is the circular array of spherical particles. Because of its symmetry, one can
expect that the highest quality factor can be
attained for particles arranged in a circle. A circle has no sharp ends
in contrast with a particle chain where the lifespan of polariton is
limited to the time it takes to travel from one end of the chain to the other \cite{Burin
LowDimArr}. It has been demonstrated \cite{Burin CircArr}\ that the bound
modes in a circular array of spherical particles possess a quality
factor that grows exponentially with the number of particles, just as in the
circular arrays of cylinder-shaped antennas (\cite%
{Fikioris,Freeman,Apalkov,Smotrova}). 

\begin{figure}[b]
\centering
\includegraphics[width=15cm]{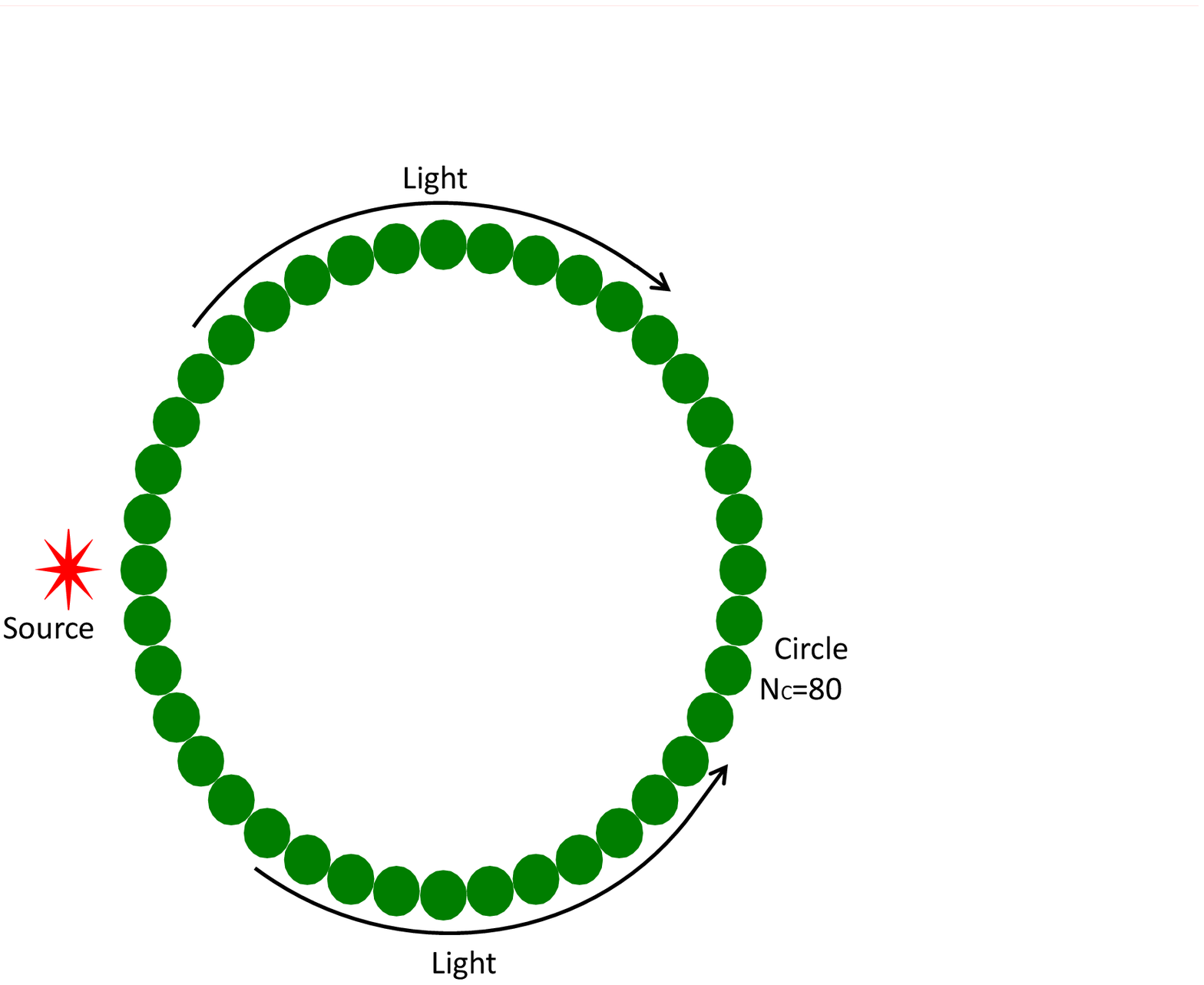}
\caption{Traffic circle without chains composed of spherical particles. 
\label{graph1A} }
\end{figure}

\begin{figure}[b]
\centering
\includegraphics[width=15cm]{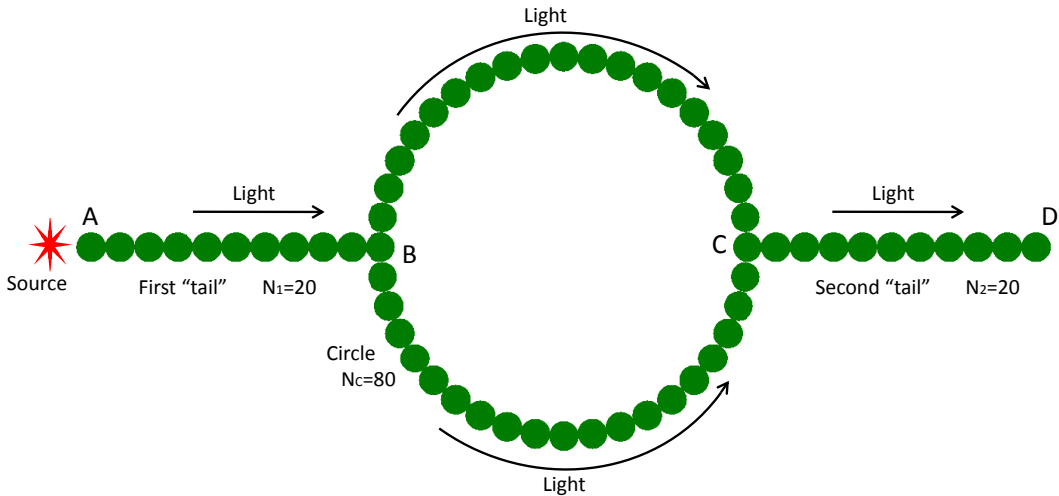}
 \caption{Traffic circle with collinear chains composed of spherical particles}
\label{graph1B}
\end{figure}

\begin{figure}[b]
\centering
\includegraphics[width=15cm]{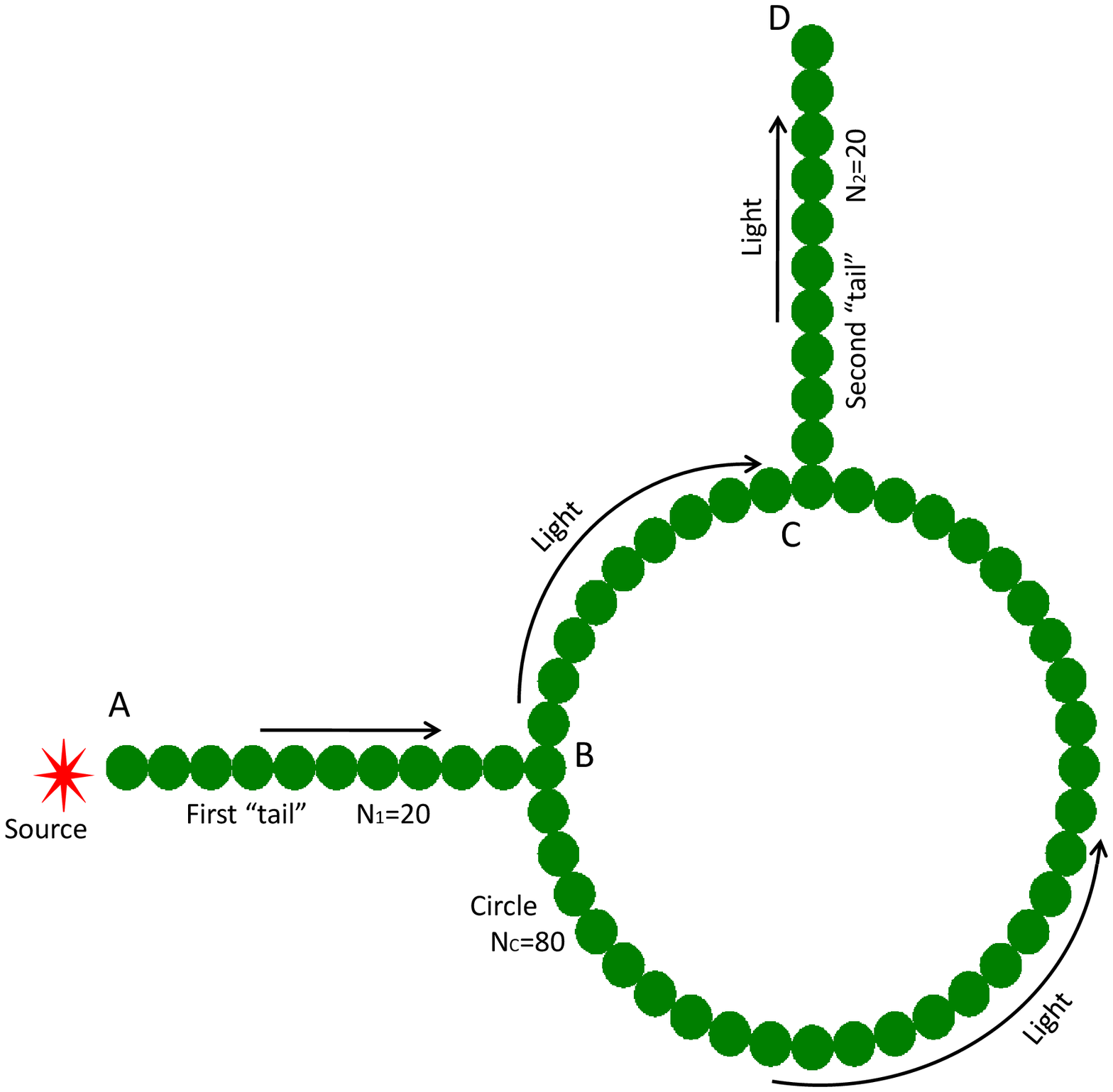}
\caption{Traffic circle with perpendicular chains composed of spherical particles}
\label{graph1C}
\end{figure}

As previously mentioned, waveguides comprised of spherical particles can be used to manipulate optical energy on a subwavelength scale. In 
addition, modes at the edge of the passband possess a vanishing group velocity which creates an interesting opportunity to realize the phenomenon of 
slow light \cite{Me OptExpress}. The properties of polariton modes within pass bands are quite similar to those of electrons within a wire. 
The question of interest is whether we can exploit several properties known about electron behavior in a wire for wave-guiding systems. In this 
manuscript, we investigate the effect of interference on the transport of optical energy. The geometry of the waveguide shown in Figs. \ref{graph1A}, 
\ref{graph1B}, \ref{graph1C} is similar to the one illustrating the Aharonov-Bohm effect in electronic systems. The sample geometry can be described 
as a "traffic circle" without an entrance or exit (Fig. \ref{graph1A}) or with them (Figs. \ref{graph1B}, \ref{graph1C}). 

The polariton is excited by a point source located near one of the aggregate particles as shown in Figs. \ref{graph1A}, \ref{graph1B}, \ref{graph1C}. 
Practically, this source can be created by
placing a dye molecule within an end particle (particle A) \cite{Astratov}. When the frequency
of the source belongs to the passband, almost all of the optical energy
generated by the source is transferred through the first chain \cite{Me PhysLett}. After
traveling through the first chain, an electromagnetic wave enters the
circle at the junction between the circle and the first chain (particle B).
It then splits into two beams, one of them propagating around the circle
clockwise and the other counterclockwise. These two beams rejoin at the
junction between the circle and the second chain (particle C) and travel
through the second chain to particle D.


Actually, the beam propagating along the circle clockwise from particle B to
particle C may pass through particle C from one segment of the circle to
the other segment and continue traveling from particle C back to
particle B. Analogously, the other beam propagating along the circle
counterclockwise from particle B to particle C also may pass through particle C and
continue traveling back to particle B in the other
segment of the circle. This recirculation in both the clockwise and counterclockwise directions remarkably enhance interference effects. 
Therefore, we investigate the interference of waves having frequencies 
corresponding to resonances of circular arrays where this enhancement is maximized. 

To describe our expectations about the interference effect, we can suggest the following simple scenario perfectly that is also 
applicable to an electronic system. 
Each of two beams of light may be expressed as a wave $\exp \{ikx\}$ if
the beam travels clockwise and $\exp \{ik(L-x)\}$ if the beam travels
counterclockwise. Here, $k$ is the wave vector of the beam and it is defined by the
frequency of the source, $x$ is the coordinate of a point on the circle (we
suppose that for point B $x=0$) and $L$ is the circumference of the circle. So, if
these two beams interfere, the intensity of the field is given by the superposition of
these two waves: 
\begin{equation}
\exp \{ikx\}+\exp \{ik(L-x)\}.
\label{1}
\end{equation}
This interference may have a number of interesting applications including, for instance, filtering and selective enhancement 
of a guiding signal. It is not clear though whether this simple interference model Eq. (\ref{1}) is applicable to optical modes because of 
the strong spatial dispersion and essential non-local coupling of spherical particles composing the waveguide. In this manuscript, we show that 
despite the above-mentioned problems, our simple interference model Eq. (\ref{1}) works quite well. 

To ascertain that interference of polaritons in a circular array takes place,
we have provided numerical investigations of a circle with and without attached chains. We have studied three particular cases: a circle without
chains, a circle with chains situated diametrically opposed to each
other and a circle with chains situated orthogonally to each other. For each
of these three cases, we have investigated two situations: the presence and the absence of absorption.

The system under consideration is a traffic circle composed of $N_{C}=80$ identical
dielectric spherical particles (see Figs. \ref{graph1A}, \ref{graph1B}, \ref{graph1C}). The
radii of particles were chosen to be unity.
The chains (linear arrays externally joined to the circle) are composed
of same particles. The number of particles
in these chains is $N_{1}=N_{2}=20$. The angle between the chains is denoted $\alpha
$. If the chains are collinear, then one has $
\alpha =0$ Fig. \ref{graph1B}, and if the chains are orthogonal to each other then one has $\alpha
=\pi /2$ Fig. \ref{graph1C}.

We introduce an oscillating magnetic dipole oriented orthogonally to the system plane as the source of electromagnetic waves. A magnetic dipole is chosen because it corresponds to the guiding band with the lowest frequency and, consequently, the best guiding properties \cite{Burin CircArr}. Indeed, as shown in the mentioned work, the lowest frequency guiding mode originates from magnetic-like Mie resonance.
If the system under consideration is a circle with chains, the source is situated near
the end of one of the chains, as  shown in Figs. \ref{graph1B}, \ref{graph1C}. If the circle
without attached chains is considered Fig. \ref{graph1A}, the source is situated outside of the circle
near the first particle.
In this manuscript, we consider GaAs particles having a refractive index 
$n_{r}=3.5$. GaAs is a convenient to consider since it possesses one of the highest refractive 
indices amongst conventional optical materials and, consequently, it has better guiding properties (cf. \cite{Burin CircArr}). The absorption effect was also probed and we discuss it when reporting the results of our calculations. 

All calculations were performed using the multisphere Mie scattering formalism (\cite{Yu-lin Xu 1,Yu-lin Xu
2}). This formalism is introduced in Sec.  \ref{sec:Mie} where we also discuss the calculations of the polariton dispersion needed to determine the resonant modes of the traffic circle. This formalism is used in Sec. \ref{sec:results} to investigate the interference of polaritons in different geometries. We investigate the intensity of the polariton around the circle in arrays shown in Figs. \ref{graph1A}, \ref{graph1B}, \ref{graph1C} as well as the intensity transferred through the waveguide with chains at arbitrary angles. The obtained dependence was compared with Eq. (\ref{1}) representing the sum of two optical beams to verify if the
dependence may be approximated with this function. 

The dependence of the intensity of an electromagnetic wave scattered by the a
spherical particle in the circle on an arbitrary point on the surface of this particle is demonstrated in Figs. \ref{graph2}, \ref{graph3}, \ref{graph4}, \ref{graph6}, \ref{graph8}. It was found that in all investigated cases the dependence may be
approximated by Eq. (\ref{1}).

In Sec. \ref{sec:conclusion} the results are summarized and their applications to various systems of interest are discussed.

\section{Multisphere Mie Scattering Formalism.}
\label{sec:Mie}

\subsection{General formalism}
 
The multisphere Mie scattering formalism \cite{Yu-lin Xu 1,Yu-lin Xu 2} (see
also \cite{Burin CircArr,Burin LowDimArr,Me ICTON2007, Me ICTON2008, Me
OptExpress, Me PhysLett}) has been developed to study the scattering of
electromagnetic waves in aggregates of spherical dielectric particles. 
The electromagnetic field outside each particle  may
be represented as the sum of all incident waves falling on this particle (say, the $j$-th particle) and of
diverging waves scattered by this particle: ${\bm E}({\bm r})={\bm E}_{inc}^{j}({\bm r})+{\bm E
}_{sca}^{j}({\bm r})$,~${\bm H}({\bm r})={\bm H}_{inc}^{j}({\bm r})+{\bm H}
_{sca}^{j}({\bm r})$. The incident wave ${\bm E}
_{inc}^{j}({\bm r})$, ${\bm H}_{inc}^{j}({\bm r})$ may be expressed as a
linear combination of spherical vector harmonics ${\bm N}_{mn}^{(1)}({\bm r}
) $, ${\bm M}_{mn}^{(1)}({\bm r})$ taken with respect to the center of $j^{th}$ particle 
\begin{eqnarray}
{\bm E}_{inc}^{j}({\bm r}) &=&\sum_{n=1}^{+\infty }\sum_{m=-n}^{n}i\left(
p_{mn}^{j}{\bm N}_{mn}^{(1)}({\bm r}-{\bm r}_{j})+q_{mn}^{j}{\bm M}%
_{mn}^{(1)}({\bm r}-{\bm r}_{j})\right) ,  \label{Einc} \\
{\bm H}_{inc}^{j}({\bm r}) &=&\sum_{n=1}^{+\infty }\sum_{m=-n}^{n}\left(
q_{mn}^{j}{\bm N}_{mn}^{(1)}({\bm r}-{\bm r}_{j})+p_{mn}^{j}{\bm M}%
_{mn}^{(1)}({\bm r}-{\bm r}_{j})\right) .  \label{Hinc}
\end{eqnarray}
Analogously, the scattered wave ${\bm E}_{sca}^{j}({\bm r})$, ${\bm H}%
_{sca}^{j}({\bm r})$ is a linear combination of spherical vector harmonics $%
{\bm N}_{mn}^{(3)}({\bm r})$, ${\bm M}_{mn}^{(3)}({\bm r})$ corresponding to outgoing waves
\begin{eqnarray}
{\bm E}_{sca}^{j}({\bm r}) &=&\sum_{n=1}^{+\infty }\sum_{m=-n}^{n}i\left(
a_{mn}^{j}{\bm N}_{mn}^{(3)}({\bm r}-{\bm r}_{j})+b_{mn}^{j}{\bm M}%
_{mn}^{(3)}({\bm r}-{\bm r}_{j})\right) ,  \label{Esca} \\
{\bm H}_{sca}^{j}({\bm r}) &=&\sum_{n=1}^{+\infty }\sum_{m=-n}^{n}\left(
b_{mn}^{j}{\bm N}_{mn}^{(3)}({\bm r}-{\bm r}_{j})+a_{mn}^{j}{\bm M}%
_{mn}^{(3)}({\bm r}-{\bm r}_{j})\right) .  \label{Hsca}
\end{eqnarray}
Here ${\bm r}_{j}$ is the coordinate of center of the $j^{th}$ particle; indices $j=1, ... N$ enumerates all particles, $N$
is the total number of particles in the array; index $n=1,...,+\infty $ represents 
angular momenta of spherical vector harmonics, and index $m=-n,...,n$ stands for the projection of angular momentum on the $z$-axis. 
The coefficients $a_{mn}^{j}$, $%
b_{mn}^{j}$, $p_{mn}^{j}$, $q_{mn}^{j}$ are partial amplitudes of scattered and incoming spherical waves at the $j^{th}$ sphere. 

The incident wave falling on the $j$-th particle is the sum of the external waves
emitted by the source and the waves scattered by all other particles.
Because of the boundary conditions at each sphere, amplitudes of these waves obey the relationship
\begin{equation}
\begin{array}{c}
\frac{a_{mn}^{j}}{\bar{a}_{n}}+\sum\limits_{l=1(l\neq
j)}^{N}\sum\limits_{\nu =1}^{+\infty }\sum\limits_{\mu =-\nu
}^{\nu }\left( A_{mn\mu \nu }^{lj}a_{\mu \nu }^{l}+B_{mn\mu \nu
}^{lj}b_{\mu \nu }^{l}\right) =p_{mn}^{j}, \\
\frac{b_{mn}^{j}}{\bar{b}_{n}}+\sum\limits_{l=1(l\neq
j)}^{N}\sum\limits_{\nu =1}^{+\infty }\sum\limits_{\mu =-\nu
}^{\nu }\left( A_{mn\mu \nu }^{lj}b_{\mu \nu }^{l}+B_{mn\mu \nu
}^{lj}a_{\mu \nu
}^{l}\right) =q_{mn}^{j}.%
\end{array}
\label{Syst}
\end{equation}
where $\bar{a}_{n}$ and $\bar{b}_{n}$ are Mie scattering coefficients 
depending on the size and the refraction index of particles. They originate from boundary conditions for an electric field between particles and air. Parameters $A_{mn\mu \nu }^{lj}$ and $B_{mn\mu \nu }^{lj}$ are vector translation
coefficients that depend on the relative position of spheres $j$ and $l$. These coefficients express delayed multipole interactions of different spheres in the frequency domain \cite{Burin CircArr}. 
These translation coefficients can be used to represent the outgoing 
wave scattered by the $l^{th}$ particle as a sum of
spherical vector harmonics ${\bm N}_{mn}^{(1)}({\bm r})$, ${\bm M}
_{mn}^{(1)}({\bm r})$ taken with respect to the $j^{th}$ particle as 
\begin{eqnarray}
{\bm M}_{\mu \nu }^{(3)}({\bm r}-{\bm r}_{l}) &=&\sum_{n=1}^{+\infty
}\sum_{m=-n}^{n}\left( A_{mn\mu \nu }^{lj}{\bm M}_{mn}^{(3)}({\bm r}-{\bm r}
_{j})+B_{mn\mu \nu }^{lj}{\bm N}_{mn}^{(3)}({\bm r}-{\bm r}_{j})\right)
,~~~~~~~~~~  \label{M3=M1,N1} \\
{\bm N}_{\mu \nu }^{(3)}({\bm r}-{\bm r}_{l}) &=&\sum_{n=1}^{+\infty
}\sum_{m=-n}^{n}\left( B_{mn\mu \nu }^{lj}{\bm M}_{mn}^{(3)}({\bm r}-{\bm r}
_{j})+A_{mn\mu \nu }^{lj}{\bm N}_{mn}^{(3)}({\bm r}-{\bm r}_{j})\right)
.~~~~~~~~~~  \label{N3=M1,N1}
\end{eqnarray}

To complete the setup of the problem Eq. (\ref{Syst}), one has to determine the coefficients $p_{mn}^{j}$ and  $p_{mn}^{j}$ for the incoming wave of the point magnetic dipole source having the coordinate ${\bm r}_{0}$ (see Figs. \ref{graph1A}, \ref{graph1B}, \ref{graph1C}). 
This external wave can be expressed as a linear combination of spherical vector harmonics $
{\bm N}_{mn}^{(3)}({\bm r})$, ${\bm M}_{mn}^{(3)}({\bm r})$ 
\begin{eqnarray}
{\bm E}_{extr}({\bm r}) &=&\sum_{n=1}^{+\infty }\sum_{m=-n}^{n}i\left( P_{mn}%
{\bm N}_{mn}^{(3)}({\bm r}-{\bm r}_{0})+Q_{mn}{\bm M}_{mn}^{(3)}({\bm r}-%
{\bm r}_{0})\right) ,  \label{Eextr} \\
{\bm H}_{extr}({\bm r}) &=&\sum_{n=1}^{+\infty }\sum_{m=-n}^{n}\left( Q_{mn}%
{\bm N}_{mn}^{(3)}({\bm r}-{\bm r}_{0})+P_{mn}{\bm M}_{mn}^{(3)}({\bm r}-%
{\bm r}_{0})\right) .  \label{Hextr}
\end{eqnarray}

Here $P_{mn}$ and $Q_{mn}$ are partial amplitudes of an emitted wave
characterizing the point source. As mentioned above, we chose the source to be the magnetic dipole parallel to the $z$ direction and perpendicular to the $x-y$ sample plane.  Then the wave emitted by the source can be 
described by the parameters $P_{mn}$ and $Q_{mn}$ defined as  $%
P_{mn}=Q_{mn}=0$ for $n>1$, also$P_{-11}=P_{01}=P_{11}=0$, $%
Q_{-11}=Q_{11}=0$, and $Q_{01}=1$.

The coefficients $p_{mn}^{j}$ and $q_{mn}^{j}$ are associated with coefficients $P_{mn}$
and $Q_{mn}$ by means of the relations 
\begin{eqnarray}
p_{mn}^{j} &=&\sum_{\nu =1}^{+\infty }\sum_{\mu =-\nu }^{\nu }\left(
A_{mn\mu \nu }^{j0}P_{\mu \nu }+B_{mn\mu \nu }^{j0}Q_{\mu \nu }\right) ,
\label{pj=PQ} \\
q_{mn}^{j} &=&\sum_{\nu =1}^{+\infty }\sum_{\mu =-\nu }^{\nu }\left(
B_{mn\mu \nu }^{j0}P_{\mu \nu }+A_{mn\mu \nu }^{j0}Q_{\mu \nu }\right),
\label{qj=PQ}
\end{eqnarray}
where vector translation coefficients are taken between the source and the center of the corresponding sphere. 
Thus, using Eqs.(\ref{pj=PQ}) and (\ref{qj=PQ}) one can calculate the
right hand side of Eq.(\ref{Syst}).

Eq.(\ref{Syst}) has the form of a system of linear equations, with partial
amplitudes $a_{mn}^{j}$ and $b_{mn}^{j}$ yet to be determined. Our main task is 
to calculate these partial amplitudes and to investigate their dependence on
the number of spheres.

In this work, the dipolar approximation is used which is sufficiently accurate to characterize the system \cite{Burin CircArr,Burin LowDimArr,Me
OptExpress}. Using this approximation, Eq.(\ref{Syst}) can be expressed as 
\begin{equation}
\begin{array}{c}
\frac{a_{m1}^{j}}{\bar{a}_{1}}+\sum\limits_{l=1(l\neq
j)}^{N}\sum\limits_{\mu =-1}^{1}\left( A_{m1\mu 1}^{lj}a_{\mu
1}^{l}+B_{m1\mu 1}^{lj}b_{\mu 1}^{l}\right) =B_{m1}^{0j}, \\
\frac{b_{m1}^{j}}{\bar{b}_{1}}+\sum\limits_{l=1(l\neq
j)}^{N}\sum\limits_{\mu =-1}^{1}\left( A_{m1\mu 1}^{lj}b_{\mu
1}^{l}+B_{m1\mu 1}^{lj}a_{\mu 1}^{l}\right) =A_{m1}^{0j}. 
\end{array}
\label{Syst1}
\end{equation}  
The numerical solution of these equations is used later in Sec. \ref{sec:results} to investigate interference of waves in the circular traffic waveguide Figs. \ref{graph1A}, \ref{graph1B}, \ref{graph1C}.

\subsection{Determination of the Dispersion Law}
\label{sec:dispersion}
 
Partial amplitudes $a_{mn}^{j}$ and $b_{mn}^{j}$
characterize interference of two polaritons running clockwise and counterclockwise. The parameter $k$
entering into Eq.(\ref{1}) is the quasi-wave vector of a polariton and it is
determined by the frequency $\omega$ of the wave emitted by
the source. The relationship between $\omega $ and $k$ is the wave 
dispersion law which has to be determined for the analysis of our simulation data. Below we briefly described how we did that. 

If the radius of the circle is large enough as in our case (a circle containing $N=80$ particles), one can use the dispersion law for an infinite linear chain. Any frequency $\omega $
lying in the conducting band of an infinite linear chain corresponds to a
certain value of the quasi-wave vector $k$. 

The frequency domain of bound modes, or the pass band, is the domain of
frequencies where losses due to photon emission are completely
suppressed in the limit $N\rightarrow \infty$. This frequency domain exists where photon emission is forbidden
by the momentum conservation law. To illustrate, consider a polariton 
within an infinite chain of particles parallel to the $z$-axis. According to the
light dispersion law, we can define the wave vector of light in vacuum as $
q=\omega c$, where $c$ is the speed of light. If an optical mode can exist
outside of the chain as a free photon it still has the fixed
projection of the wave vector to the $z$-axis given by $k$ so we should have $
q_{z}=k$ due to translational symmetry. Then the perpendicular component of the electric field wave vector
that describes the photon emission outside of the system can be expressed as $
q_{\perp }=\sqrt{q^{2}-k^{2}}$. For unbound modes, this projection is real,
meaning that the optical energy can escape from the chain of particles,
while for bound modes it is imaginary so that the mode intensity decays at distance $\rho $ from the chain as $e^{-\rho \sqrt{k^{2}-q^{2}}}$. Thus modes with $\omega /c<k$ are bound, while the modes with $\omega /c>k$
are unbound. Besides that, according to the Bloch theorem the absolute
value of $k$ cannot exceed the value $\pi /a$, where $a$ is the period of the
chain. Therefore, bound modes can possess quasi-wave vectors $k$ from the
interval
\begin{equation}
\frac{\omega }{c}<k<\frac{\pi }{a}.  \label{CondBand}
\end{equation}
Certainly, this condition may be satisfied only if $\omega /c<\pi /a$. This
guiding mode criterion is equivalent to the condition that the interparticle
distance $a$ should be less than half the resonant wavelength: $a<\lambda
/2 $ \cite{Ehrespeck}. 

To find the dispersion law for a polariton within an infinite chain \cite%
{Deych,Shore,Fan,Meade}\ one has to calculate the eigenvalues of the frequency $%
\omega $ for the given quasi-wave vector $k$. The eigenfrequencies can be
computed using the condition of the existence of nontrivial solutions of
a homogeneous system of equations
\begin{equation}
\begin{array}{c}
\frac{a_{mn}^{j}}{\bar{a}_{n}}+\sum\limits_{l=1(l\neq
j)}^{N}\sum\limits_{\nu =1}^{+\infty }\sum\limits_{\mu =-\nu
}^{\nu }\left( A_{mn\mu \nu }^{lj}a_{\mu \nu }^{l}+B_{mn\mu \nu
}^{lj}b_{\mu \nu }^{l}\right) =0, \\
\frac{b_{mn}^{j}}{\bar{b}_{n}}+\sum\limits_{l=1(l\neq
j)}^{N}\sum\limits_{\nu =1}^{+\infty }\sum\limits_{\mu =-\nu
}^{\nu }\left( A_{mn\mu \nu }^{lj}b_{\mu \nu }^{l}+B_{mn\mu \nu
}^{lj}a_{\mu \nu
}^{l}\right) =0.
\end{array}
\label{Syst=0}
\end{equation}
According to the Bloch's theorem, the partial amplitude in an infinite
periodic chain can be represented by
\begin{equation}
a_{mn}^{j}=a_{mn}e^{ikx},\ \ \ b_{mn}^{j}=b_{mn}e^{ikx},  \label{Fourier}
\end{equation}
where $k$ is the quasi-wave vector and $x=aj$ is the coordinate of the center of $j^{th}$ particle in the chain (remember, we set $a=1$). 

Substituting Eq. (\ref{Fourier}) into Eq. (\ref{Syst=0}), we get (cf. Refs. \cite{Shore,Burin CircArr}) 
\begin{equation}
\begin{array}{c}
\frac{a_{mn}}{\bar{a}_{n}(\omega )}+\sum\limits_{\nu =m}^{+\infty
}\sum\left( A_{mn m \nu }(\omega
,k)~a_{m \nu }+B_{mn  m\nu }(\omega ,k)~b_{m \nu }\right) =0,
\\ \frac{b_{mn}}{\bar{b}_{n}(\omega )}+\sum\limits_{\nu
=1}^{+\infty }\left( A_{mn m\nu
}(\omega ,k)~b_{m \nu
}+B_{mn m \nu }(\omega ,k)~a_{m\nu }\right) =0.%
\end{array}
\label{FourSyst=0}
\end{equation}
We used the momentum projection $m$ conservation law due to the rotational invariance of the system.

In the dipolar approximation, Eq.(\ref{FourSyst=0}) takes the form
\begin{equation}
\begin{array}{c}
\frac{a_{m1}}{\bar{a}_{1}(\omega )}+\sum\limits_{\nu =1}^{+\infty
}\left( A_{m1m1}(\omega ,k)~a_{m1}+B_{m1m1}(\omega
,k)~b_{m1}\right) =0, \\ \frac{b_{m1}}{\bar{b}_{1}(\omega
)}+\sum\limits_{\nu =1}^{+\infty }\left(
A_{m1m1}(\omega ,k)~b_{m1}+B_{m1m1}(\omega ,k)~a_{m1}\right) =0.%
\end{array}
\label{Dip_FourSyst_m=0}
\end{equation}
If $m=\pm 1$, the modes are called $t$-modes (meaning transverse modes with polarization perpendicular to the chain), and if $m=0$ the modes are called $l$-modes (meaning longitudinal modes with polarization parallel to the chain). In this manuscript we are interested in $t$-modes. 

As was discussed above, the source of electromagnetic waves in our problem of interest is a magnetic dipole directed perpendicular to the plane of the 
circle. This source excites only transverse electromagnetic waves. Therefore we are interested in equations
for $m=1$ 
\begin{equation}
\begin{array}{c}
\frac{a_{11}}{\bar{a}_{1}(\omega )}+\sum\limits_{\nu =1}^{+\infty
}\left( A_{1111}(\omega ,k)~a_{11}+B_{1111}(\omega
,k)~b_{11}\right) =0, \\ \frac{b_{11}}{\bar{b}_{1}(\omega
)}+\sum\limits_{\nu =1}^{+\infty }\left(
A_{1111}(\omega ,k)~b_{11}+B_{1111}(\omega ,k)~a_{11}\right) =0.%
\end{array}
\label{Dip_FourSyst_(m=1)=0}
\end{equation}
This system contains only two variables $a_{11}$ and $b_{11}$
and the dispersion law is defined by the following condition 
\begin{equation}
\left(\frac{1}{\bar{a}_{1}(\omega )}+A_{1111}(\omega
,k)\right)\left(\frac{1}{\bar{a}_{2}(\omega )}+A_{1111}(\omega
,k)\right)-(B_{1111}(\omega
,k))^2=0.
\label{Determ}
\end{equation}
Eq.(\ref{Determ}) can be solved only numerically. We used the Newton-Raphson method to find quasiwavevector for certain frequencies of interest following Ref. \cite{Burin CircArr}. 

\section{Results}
\label{sec:results}

In this work three arrays of particles were considered, including the circle of
particles without attached chains Fig. \ref{graph1A} and two structures with chains Figs. \ref{graph1B} and \ref{graph1C}.
Each of these structures is investigated for particles of ideal
non-absorbing material. Our preliminary study of wave transmission through a traffic circle at various frequencies has demonstrated that it is most efficient for modes corresponding to resonances of a circular array (see Ref. \cite{Burin CircArr}). These resonances takes place at frequencies corresponding to integer projections $q=-N/2,...,N/2$ of a polariton's angular momentum to the $z$-axis. These modes can be characterized by quantized quasi-wavevectors $k=2\pi q/L$. The upper boundary of the pass
band corresponds to $q=N/2$, i.e. $k=\pi$. For the circular array under
consideration this upper boundary is given by $\omega =0.8434$. 

In this work we investigate a circular array composed by $N=80$ particles. 
To characterize the interference, we chose two modes having angular momenta 
$q_{1}=38$ and $q_{2}=37$  and frequencies $\omega _{1}=0.8424$ and $\omega _{2}=0.8413$, respectively, 
located at the top of the passband ($q=40$). Interference behavior is quite similar for other modes with $q\geq 30$. The mode at the top of the band characterized by $q=40$ does not show any interference because the wave intensity does not change from sphere to sphere; though, the field changes its sign between neighboring particles. 

The intensity of a polariton mode can be characterized by squared partial amplitudes $a_{mn}^{j}$ and $b_{mn}^{j}$. Since we are using the dipolar approximation, only coefficients $a_{mn}^{j}$ and $b_{mn}^{j}$ for $n=1$ should be considered. The partial amplitudes $a_{01}^{j}$, $%
b_{-11}^{j}$ and $b_{11}^{j}$ are all equal to zero because of the planar structure of our system. Therefore, the intensity of a
polariton within the $j$-th particle is characterized 
by three coefficients $a_{-11}^{j}$, $a_{11}^{j}$ and $b_{01}^{j}$. For the sake of simplicity, we define it as $I_{j}=\mid b_{01}^{j}\mid^2$. 
Using the multisphere Mie scattering formalism \cite{Yu-lin Xu 1}, we investigated the
dependence of intensity of the wave on the $j$-th particle on the position of the particle. The position of the particle is defined by its 
number $j$ and may be described as the length of the segment of the circle between the
first particle and the $j$-th particle counted counterclockwise: $x_{j}=(j-1)$.  

The intensity of the wave should be compared with the one following from  Eq. (\ref{1}) 
\begin{equation}
I_{j}\propto \mid \exp \{ikx\}+\exp \{ik(L-x)\}\mid^{2}/4=\cos^{2}(\pi q - kx)=\cos^{2}(2\pi q x/N).
\label{eq:I1}
\end{equation}
Since for guiding modes of interest one has $q\approx N/2$, this function strongly oscillates for non-integer $x$ and is difficult to show in our graphs. Therefore, we compare it directly to our simulations in Fig. \ref{graph2} only, while in all other graphs it is replaced with the function 
\begin{equation}
I_{j}\propto \cos^{2}(2\pi (N/2-q) x/N),
\label{eq:I2}
\end{equation}
which coincides with Eq. (\ref{eq:I1}) for all integer points $x$ of interest. For chosen modes with $q=38$ and $q=37$, this function can be expressed as $\cos^{2}(2\cdot 2\pi x/40)$ and $\cos^{2}(3\cdot 2\pi x/40)$, respectively.

Below we report our calculations for the electromagnetic field within dielectric spheres for two selected modes for all three configurations shown in Figs. \ref{graph1A}, \ref{graph1B}, \ref{graph1C} and then investigate the effect of interference on the transmitted intensity. 

\subsection{Circle without attached chains}

\begin{figure}[b]
\centering
\includegraphics[width=15cm]{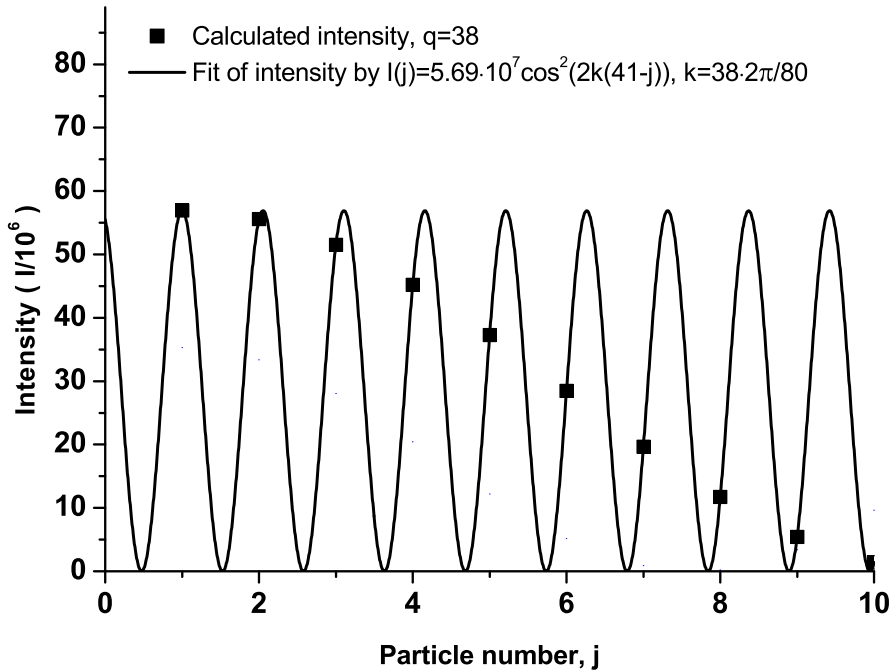}
 \caption{High resolution representation of intensity of the wave depending on particle number $j$ within a circle without attached chains}
\label{graph2}
\end{figure}

\begin{figure}[b]
\centering
\includegraphics[width=15cm]{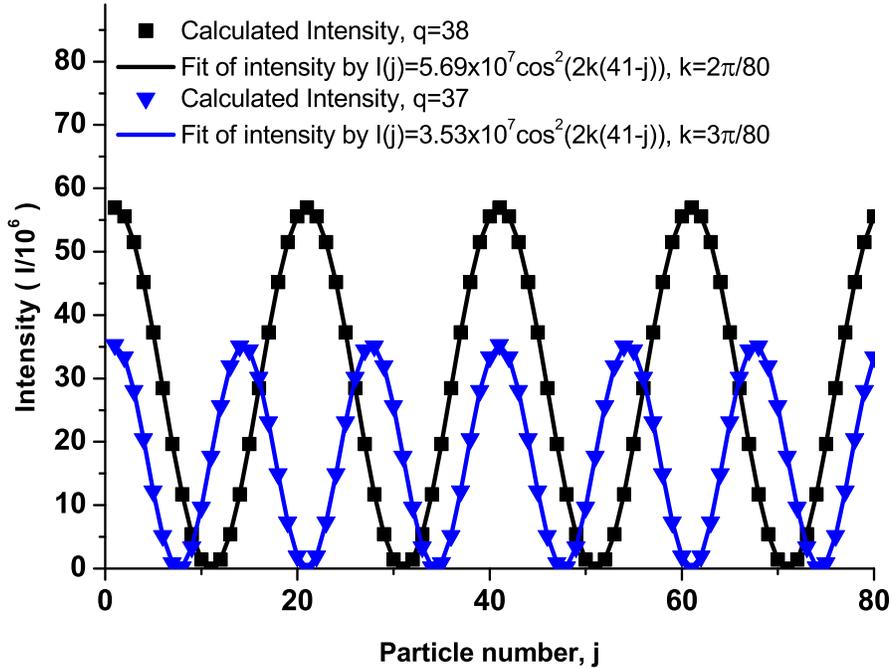}
 \caption{Intensity of a wave depending on particle number $j$ within a circle without attached chains}
\label{graph3}
\end{figure}

Consider the intensity behavior in a circle without attached chains (Fig. \ref{graph1A}). 
In Fig. \ref{graph2} and Fig. \ref{graph3}, we show the results of numerical calculations of
the wave intensity $I_{j}$ depending on the number of particles for two different resonant source frequencies $\omega
_{1}$ and $\omega _{2}$ defined previously. 
Solid curves on Figs. \ref{graph2}, \ref{graph3} represent the data fit by functions Eqs. (\ref{eq:I1}) and (\ref{eq:I2}), respectively, with a numerical prefactor chosen as a scaling factor to make the fit intensity identical to the calculated one for $j=1$.   
According to Figs. \ref{graph2} and \ref{graph3} at both frequencies $\omega
_{1}$ and $\omega _{2}$, there is a perfect agreement between our calculations and our simple interference model Eqs. (\ref{eq:I1}), (\ref{eq:I2}). 
Thus guiding modes in a
circular array excited by a point source can be treated using the 
interference of two optical beams propagating along a circle in opposite
directions.

We verified that the weak absorption  Im $n_{r} < 0.01$ does not qualitatively affect the interference behavior Eqs. (\ref{eq:I2}), but remarkably  reduces the wave intensity. This is because guiding modes of circular arrays possess extremely high quality factors \cite{Burin CircArr} growing exponentially with the number of particles. Even absorption as weak as Im $n_{r} \sim 10^{-5}$ results in remarkable broadening of all resonances.

\subsection{Traffic circle with attached chains}

Consider the traffic circle with attached chains Figs. \ref{graph1B}, \ref{graph1C}. The circle is composed of $%
N_{C}=80$ particles, and the chains are composed of $N_{1}=N_{2}=20$
particles. The source of electromagnetic waves is again a magnetic dipole
oscillating normally to the plane of the circle. It is located 
close to the end of one of the chains (see Fig. \ref{graph1B}, \ref{graph1C}).

As in the previous case, we investigate the dependence of intensity of a polariton at particle $j$ on the number $j$.  The
influence of the chains on this dependence is examined. 
We begin our consideration with the system with the chains situated
collinearly as in Fig. \ref{graph1B}.

\begin{figure}[b]
\centering
\includegraphics[width=15cm]{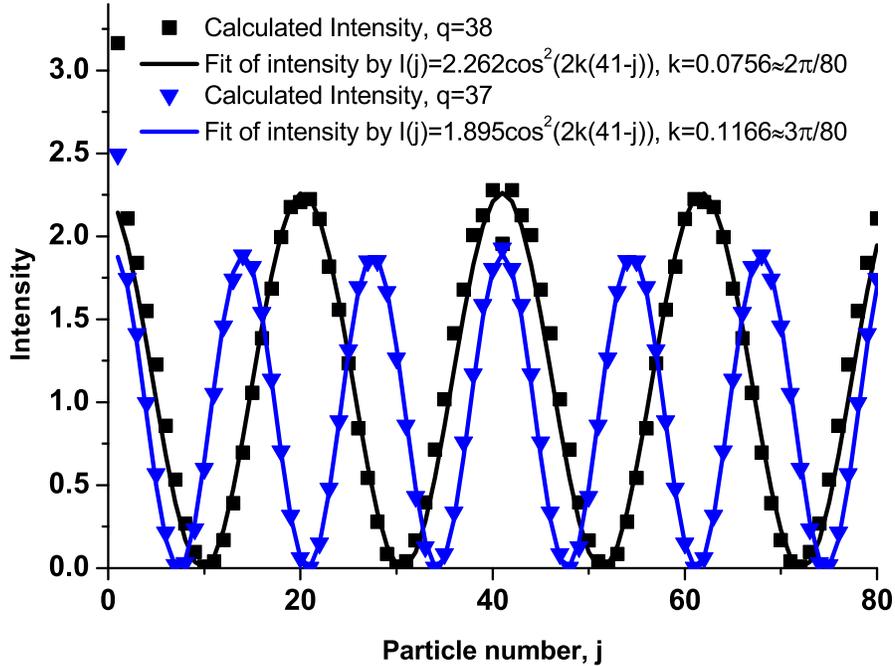}
 \caption{Intensity $I_{j}$ versus particle number $j$ within a circle with collinear chains}
\label{graph4}
\end{figure}




The results of calculation are presented in Fig. \ref{graph4}. Symbols
denote the intensity $I_{j}$ calculated
numerically, and solid lines show fit using the 
analytical function Eq. (\ref{eq:I2}) with arguments $
x_{j}=2(j-1)$. The squares and the upper solid line correspond to the frequency $\omega
_{1}=0.8424$ and the triangles and the lower solid line correspond to the frequency $\omega
_{2}=0.8413$.

It is clear that the symbols are located near the solid lines. Therefore, the dependence of intensity $I_{j}$ $j$ can be approximately expressed by the function Eq. (\ref{eq:I2}), similarly to
the traffic circle without attached chains. However, the agreement of the numerically calculated intensity with
the analytical expression Eq. (\ref{eq:I2}) is less accurate than in the case of the circle
without attached chains. This is clear because guiding mode propagation in a circle with
chains is much more complicated than in the circle without chains due to scattering at its junctions.


\begin{figure}[b]
\centering
\includegraphics[width=15cm]{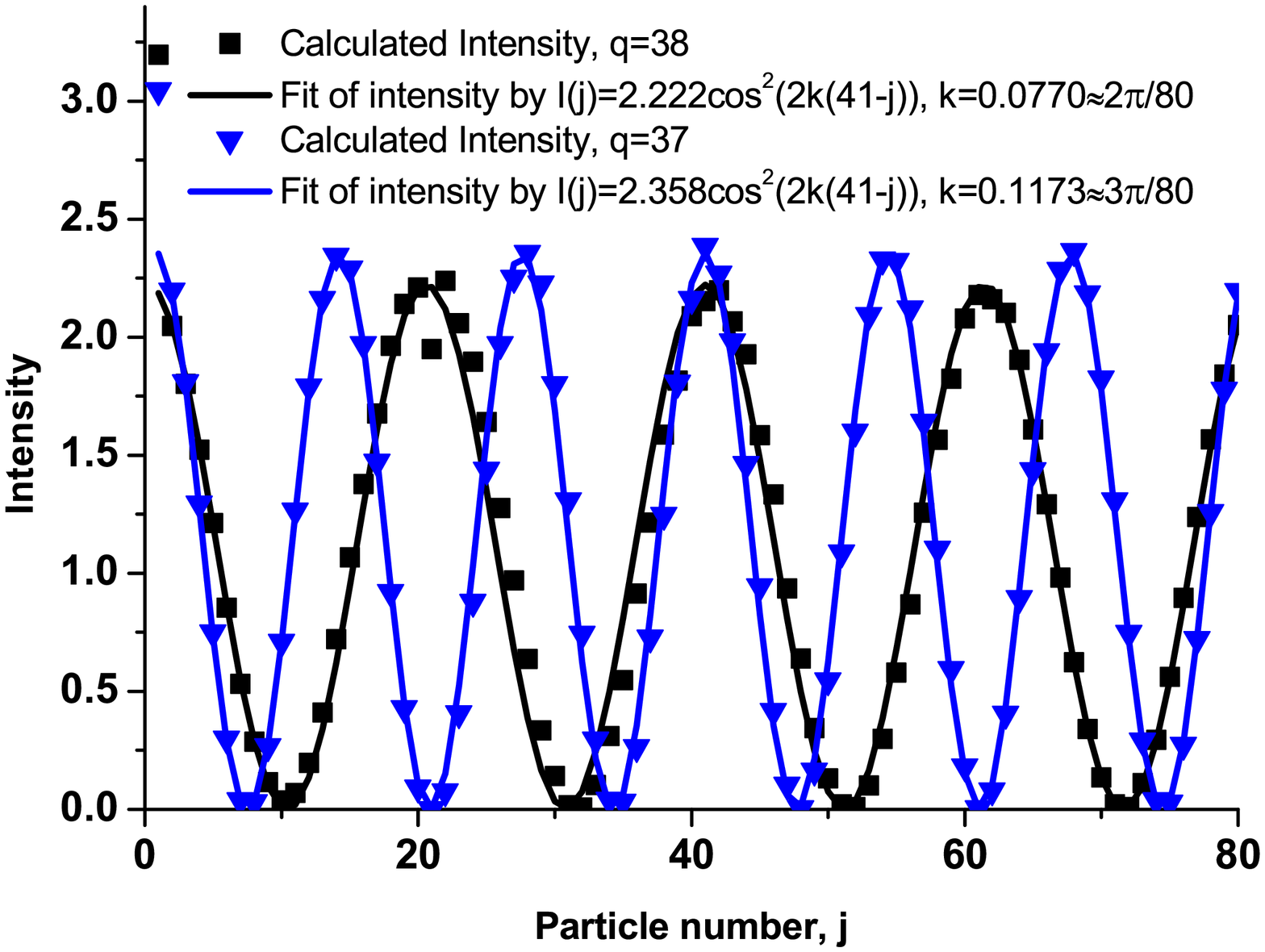}
 \caption{Intensity $I_{j}$ versus particle number $j$ within a circle with perpendicular chains}
\label{graph6}
\end{figure}



Consider a circle with chains forming right angles with each
other. In Fig. \ref{graph6}, the results of numerical calculations of the 
absolute value of the intensity $I_{j}$ are shown for the two modes under consideration.
In this case, similarly to the case of collinear
chains, the dependence of intensity on the position of a particle
in the circle may be well approximated by the function Eq. (\ref{eq:I2}). Therefore, the
approximation of intensity  by the superposition of two waves is acceptable
for any position of the second chain.

The intensities for a circle without chains and for
a circle with chains in the absence of absorption differ by six orders of magnitude. 
This effect is due to the reduction of the mode quality factor due to polariton scattering by the chains.  Indeed, quality factors 
of resonant modes of circular arrays increase exponentially with the number of particles in the circle.  
However, quality factors of circles with chains behave similarly to those of finite linear chains where they increase with 
the number of particles according to the power law \cite{Me OptExpress}. Six orders of magnitude 
difference in intensity is also due to a difference in quality factors of resonant modes under consideration.

Another interesting phenomenon is that absorption in a circle with
chains has a weaker influence on intensity. This is another consequence of the difference in quality factors. 

\subsection{Transmitted intensity affected by interference}

Traffic circle interference is characterized by its transmitted intensity. This is the intensity 
at the end of the second chain (particle D - see
Figs. \ref{graph1B}, \ref{graph1C}). We investigate its dependence on junction position, which can be any particle in the circular array. 
Similarly to Figs. \ref{graph1B}, \ref{graph1C} the second chain is placed in a radial direction perpendicular to the circle, 
while its starting point can any arbitrary particle within the circle. This dependence characterizes the interference effect on the output signal. 

We expect that the dependence can still  be approximated (at least roughly) by
the function Eq. (\ref{eq:I2}) assuming that the intensity at the end of the second chain should be proportional to the
intensity of field at its origin. Since the intensity in the $j$-th particle is defined by the interference of polaritons and may be approximated by Eq. (\ref{eq:I2}), one can assume that the intensity at the end of the chain behaves similarly.

\begin{figure}[b]
\centering
\includegraphics[width=15cm]{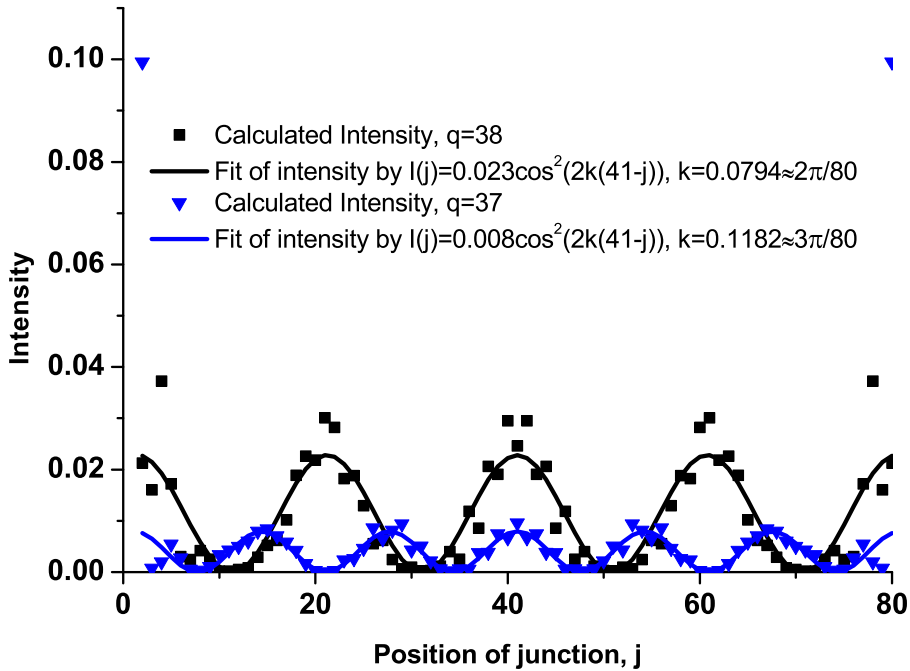}
 \caption{Dependence of transmitted intensity on the position of junction.}
\label{graph8}
\end{figure}

In Fig. \ref{graph8}, the calculated dependence of the transmitted intensity $I_{j}$ on the junction position $j$ is shown. As shown previously, the intensities calculated numerically are marked by symbols and analytical fits by Eq. (\ref{eq:I2}) are shown using solid lines. One can see that the dependence of partial amplitude $b_{01}$ at the end of
the second chain on the junction position can be approximated by
the function Eq. (\ref{eq:I2}). Yet this approximation is rough. For several positions of the junction the error can be as high as  
$30\%$. Still this approximation is convenient for a qualitative description of behavior of an
electromagnetic field at the end of the second chain.
The influence of absorption on the transmitted intensity of field
at the end of the second chain is weak similarly to a circle without chains as considered previously.


\section{Conclusion}
\label{sec:conclusion}

In this work we have investigated the interference of guiding modes in
a traffic circle waveguide of dielectric spherical particles using the multisphere Mie scattering formalism at wave frequencies corresponding to two resonant guiding modes.
For different sample geometries we demonstrated that the field distribution within the circle as well as the transmitted intensity may be well approximated by interference of two optical beams propagating clockwise and counterclockwise around
the circle.  

The main conclusion of this work is that strong interference exists in the waveguides composed of particles and is quite similar to interfenrence as seen in electronic systems. This interference can be used to control the output (transmitted) intensity through varying the system geometry and/or the mode frequency. These results must also be applicable to the subwavelength waveguides consisting of metal particles \cite{Mayer1,Mayer2} despite absorption by conducting electrons and can be used to narrow the transferring signal in the frequency domain.  

It is difficult to apply our results directly to experiments with whispering gallery modes in arrays of polystyrene particles \cite{Astratov} because the dipolar approach is irrelevant for these modes having large angular momenta. However, interference effects might take place there because coupling of spheres is essentially local \cite{Deych}. Another interesting application can be developed for the propagation of microwaves in arrays of coupled antennas \cite{Shore,Fikioris} which are very similar to our particle aggregates. Interference can be used in this application for frequency selective emission, absorption or transport of the microwave signal. 

We do not discuss in detail frequency dependence of the transmitted intensity. Our preliminary study shows that this dependence is very complicated because the interference of clockwise and counter clockwise beams is superimposed with circular array resonances. This problem is still under investigation and we do not have a reasonable interpretation for it yet.

This work is supported by the Air Force Office of Scientific Research (Award no. FA 9550-06-1-0110). Authors acknowledge fruitful discussions with Vasilii Asratov, Olga Samoylova, Mark Sulkes and Arthur Yahjian.

\end{document}